\documentclass[12pt,preprint]{aastex}
\usepackage{graphicx}
\usepackage{amssymb}
\usepackage{natbib}
\usepackage{longtable,lscape}
\usepackage{subfigure}

\begin{document}

\title{New Members of the Scorpius Centaurus Complex and Ages
       of its sub-regions}

\author{Inseok Song\altaffilmark{1},
        B.~Zuckerman\altaffilmark{2},
        M.~S.~Bessell\altaffilmark{3}}

\altaffiltext{1}{ Department of Physics and Astronomy,
              The University of Georgia,
              Athens, GA 30602--2451, USA}
\altaffiltext{2}{ Dept. of Physics \& Astronomy,
              University of California, Los Angeles,
              475 Portola Plaza,
              Los Angeles, CA 90095--1547, USA}
\altaffiltext{3}{ Research School of Astronomy and Astrophysics,
              Institute of Advanced Studies,
              The Australian National University, ACT 2611, Australia}

\begin{abstract}
We have spectroscopically identified $\sim\!\!100$ G-, K- and M-type
members of the Scorpius Centaurus complex.  To deduce the age of these
young stars we compare their Li\,$\lambda6708$ absorption line strengths
against those of stars in the TW~Hydrae association and $\beta$~Pictoris
moving group.  These line strengths indicate that ScoCen stars are
younger than $\beta$~Pic stars whose ages of $\sim$12 Myr have
previously been derived from a kinematic traceback analysis.  Our
derived age, $\sim\!\!10$\,Myr, for stars in the LCC and UCL subgroups
of ScoCen is younger than previously published ages based on the moving
cluster method and upper main sequence fitting. The discrepant ages are
likely due to an incorrect (or lack of) cross-calibration between
model-dependent and model-independent age-dating methods.
\end{abstract}
\maketitle

\section{Introduction}

The Scorpius-Centaurus region (ScoCen) is the nearest ($100-200$\,pc)
massive star formation site to Earth. It consists of three subgroups
\citep{deZeeuw}; Upper Scorpius (US), Upper Centaurus Lupus (UCL), and
Lower Centaurus Crux (LCC). Each of these sub-regions has a different
location in the sky plane, different age, and different space motion.
Therefore, ScoCen is the best site for studying a sequential star
formation or triggered star formation phenomena. Furthermore, ScoCen
holds the key to the origin of nearby young stellar groups
\citep{ARAA,Fernandez}, and due to its youth and proximity to Earth a thorough
membership determination can be made down to very low mass.

Using Hipparcos data, \citet{deZeeuw} refined a list of ScoCen
members containing many B- and A-type stars and relatively few F-
and G-type stars. From the number of B and A-type stars
(N$\sim\!\!300$), several thousand low mass members were expected
to exist.  However, mainly due to the vast surface area of
ScoCen in the projected sky plane ($\sim\!2000$\,deg$^2$) and its
deep southern declination (80\,\% of the region is below
dec=$-40$\,deg), this region has been little studied.  Compared to
a similarly massive but more distant star formation site (e.g.,
the Orion region), the Sco-Cen complex has been barely
investigated. No systematic search for low mass members of ScoCen
has been carried out with the exception of occasional small area
pilot surveys \citep[for example][]{Preibisch}. 

\citet{Mamajek} identified several dozen F- and G-type LCC and UCL
members. Using a moving cluster method, they estimated secular
parallaxes of new members, then derived ages (UCL 16\,Myr and LCC
17\,Myr) by plotting them on a Hertzsprung-Russell diagram and comparing
them with theoretical pre-main sequence models. Recently, using a larger
($N$=138) sample of F-type kinematic members from \citet{deZeeuw},  \citet{Pecaut}
re-deduce old ages (16/17 Myr) for the UCL/LCC regions. 

Using our $\sim\!\!100$
spectroscopically confirmed G/K/M-type members of LCC/UCL (Table 1), we show that
the LCC/UCL age is more consistent with a younger age
($\sim\!\!10$\,Myr).  Because our age determination is anchored in the
traceback age for the $\beta$~Pic moving group and because kinematic
traceback is the least model dependent technique for deriving ages of
young stars, we expect that a $\sim$10 Myr age for LCC/UCL is most
likely to be correct.

Ages of LCC and UCL are important in the interpretation of Spitzer
and other data (e.g., Currie et al 2008). For example, the high
fraction ($>35$\,\%) of dusty disks (including several mid-IR
excesses) around Sco-Cen F/G stars \citep{Chen} applies to
$\sim$10\,Myr old stars, rather than stars of nearly twice this age.

\section{New Members}

\subsection{Observations}

As part of an extensive search for young and nearby stars to Earth,
$\lambda/\Delta\lambda\sim4,500$ spectra of candidate ScoCen members
were obtained with the Double Beam Spectrograph (DBS) on the Nasmyth-A
focus of the Australian National University's 2.3\,m telescope.  For
many bright young stars confirmed from DBS spectra, we later obtained
echelle spectra to obtain radial velocities.  Candidate ScoCen members
were selected over the ScoCen region (Figure~9 of \citealt{deZeeuw})
from a correlation between X-ray (ROSAT; \citealt{RASS,RASSFSC}) and
kinematic catalogs (Hipparcos: \citealt{HIP}, Tycho-2: \citealt{TYC2},
SuperCOSMOS: \citealt{SSS}).  Then, we kept only X-ray bright stars
($\log L_X/L_{bol}\gtrsim\!10^{-3.5}$) whose space motions are
consistent with the nominal value of LCC ($UVW=-12,-13,-7$ km/sec;
\citealt{deZeeuw}, details on space motion calculation are given
below).  Since most candidate members lack sufficient information to
enable direct calculation of their $UVW$ (distances to non-Hipparcos stars
and radial velocities for almost all candidate members), we calculated
$UVW$ based on photometric distances using an $\sim$10\,Myr age and a
range of radial velocities ($-50$ to $+50$\,km/sec). If an X-ray star
can have a ScoCen-like $UVW$ for some radial velocities within the
above stated range, then we selected it as a candidate ScoCen member.
Our chosen radial velocity range is large enough to cover nearly all Sco-Cen
members because a typical speed ($\equiv\sqrt{U^2+V^2+W^2}$) of ScoCen
stars is less than 30\,km/sec requiring that their radial velocities
should be smaller than this total speed. The use of young
age in the photometric distance estimate does not affect
the candidate selection much because of our rather generous $UVW$ range
($\pm5$\,km/sec in each component).

\begin{figure}
\centering
 \includegraphics[width=0.95\columnwidth]{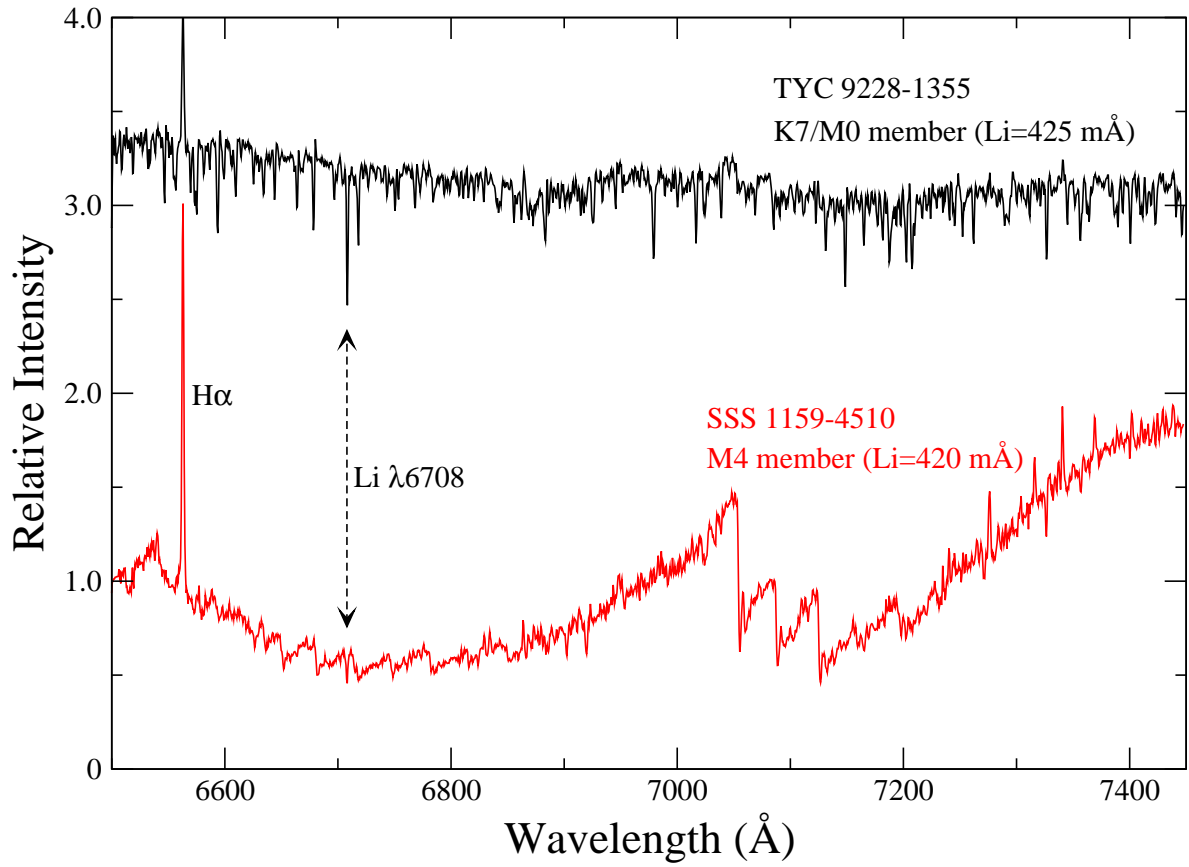}
   \caption{Representative spectra of newly identified ScoCen members.}
   \label{spectra}
\end{figure}

All spectra were reduced with IRAF following a standard procedure (bad
pixel and cosmic ray removal, flat fielding, source extraction, telluric
correction, etc.). Typical spectra have $\sim$5000 counts pixel$^{-1}$
in the vicinity of 6700\,\AA. Equivalent widths of Li\,I~$\lambda$6708
and H$\alpha$ together with their $V-K$ colors and X-ray information are
listed in Table~1.  We used $V-K$ colors (see footnotes of Table~1 for
sources of $V$ and $K$ magnitudes) as spectral type proxies because
$V-K$ separates K- and M-type subclasses nicely and the long color
baseline is less susceptible to measurement errors and time
variabilities compared to other broadband colors (e.g., $B-V$).  Two
typical spectra are displayed in Figure~\ref{spectra}.

\subsection{Refinement of LCC and UCL Ages}

To estimate ages of LCC and UCL from our spectra, in Figure 2 we compare
their Li\,$\lambda6708$ absorption strengths against those of other
young stellar groups with well known ages on a EW(Li) versus $V-K$ plot.
Because Table~1 contains US stars that are thought to be
$\sim\!\!5$\,Myr old, we plot LCC/UCL stars and US stars with different
symbols on Figure~2.  Ages of
the TW~Hydrae Association ($\sim\!\!8$\,Myr) and the $\beta$~Pictoris
Moving Group ($\sim\!\!12$\,Myr) are well established and calibrated
against contemporary theoretical pre-main sequence evolutionary
models by plotting their members on a color magnitude diagram along with
theoretical models. An essentially model-independent age of the
$\beta$~Pictoris moving group was obtained by tracing positions of its
members backward in time \citep{Ortega,Song03}.  Members of this unbound
stellar group should have been confined in the smallest volume at its
birth; a kinematic age of $\sim\!\!12$\,Myr obtained from this method
agrees well with current stellar evolution models. Similarly, a
kinematic age of TWA is estimated to be $8.3\pm0.8$\,Myr
\citep{deLaReza06}.

As is evident in Figure~2, overall lithium absorption strengths
of LCC/UCL stars fall between those of the TWA and the
$\beta$~Pictoris moving group.  Therefore, a likely age of LCC/UCL is
$\sim\!\!10$\,Myr.  Reddening toward the LCC/UCL region ($A_J = 0.00 -
0.35$\,mag, \citealt{Mamajek}) does not change the relative ordering
of Li\,$\lambda6708$ strength distribution among TWA, LCC/UCL, and
$\beta$~Pictoris Moving Group members. In fact, dereddening will make
most LCC/UCL stars appear younger (i.e., moving LCC/UCL stars leftward
in Figure 2) because reddening toward the TWA and the $\beta$~Pictoris
Moving Group is almost negligible. For this reason, we do not consider
the effect of reddening in this paper.   

Current theoretical stellar evolutionary models (e.g., \citealt{BCAH})
predict near complete depletion of lithium (down to the 1-2\,\% level)
among M1/2 stars ($V-K\sim4.0$) within 16\,Myr, but we do not see such
depletion of lithium among early M-type LCC/UCL stars (i.e.,
$V-K\sim4.0$) in Figure~2.  In addition, lithium depletion rates
predicted in current evolutionary models appear to be slower than what
is observed (e.g., \citealt{LDB}) which further strengthens the
preceding statement.  As demonstrated in Figure~2, LCC/UCL members are
younger than $\beta$~Pictoris moving group members.  Therefore, the
LCC/UCL cannot be as old as 16\,Myr.

\begin{figure}
 \resizebox{\hsize}{!}{\includegraphics{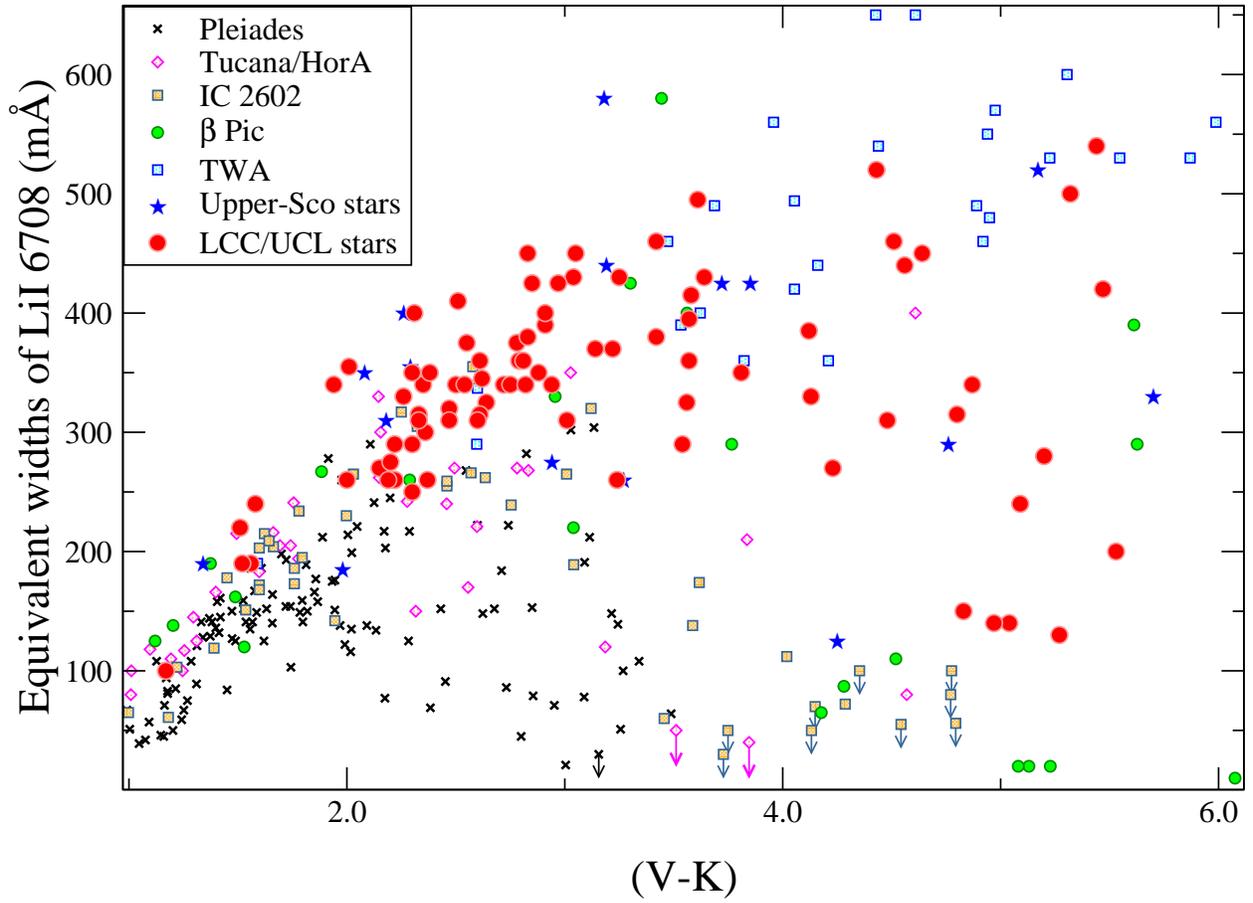}}
 \caption{Li\,$\lambda6708$ equivalent widths of LCC and UCL stars.  
Tucana/HorA and IC 2602 stars are generally considered to be $\sim$30\,Myr 
old.}
\end{figure}

Due to our target selection criterion based on ROSAT all-sky X-ray
detection, our Table~1 LCC stars are systematically closer than
stars considered by \citet{Mamajek}. Therefore, it is conceivable
that our LCC/UCL stars (close to Earth) are $\sim\!\!10$\,Myr old
while more distant LCC/UCL stars surveyed by \citet{Mamajek} could
be 17/16\,Myr old. Supporting this conjecture, although based on small number
statistics, \citet{TWA_Rot}
photometrically measure the median rotational period of TWA\,1--13
(4.7 days) to be longer than the median value of TWA\,14--19 (0.7
days) which they interpret as an age difference between these two
groups; TWA\,1--13 being younger than TWA\,14--19. To investigate
the possibility of age dependence on distance for our low mass
stars, we divided LCC/UCL stars of Table~1 into two groups
(distance $\leqq95$\,pc [N=46] and distance $>95$\,pc [N=41]) and
compared their Li absorption strengths. Stars in these two bins
show almost identical Li\,$\lambda$6708 absorption strength
distribution, hence we believe that the whole LCC/UCL group is
$\sim\!\!10$\,Myr old.  Furthermore, TWA\,14--19 all show very
strong Li\,$\lambda6708$ absorption strengths that are consistent
with LCC stars in Table~1. Nonetheless, whether there is a radial
age spread toward the direction of LCC/UCL or not, from Figure~2
alone, G/K/M-type members of LCC/UCL are younger than
$\sim$12\,Myr old. Also, since the original discovery of TWA, many 
more members have been identified to date. These newly discovered
members are generally more distant than original TWA members. As a result, the apparent clear
distinction in distance between original TWA members and LCC members 
is disappearing. The current set of age-dating methods cannot
readily discern $\sim$8\,Myr old stars from $\sim$10\,Myr ones. Based on 
several common characteristics (similar ages, similar positions on
the projected sky plane, similar space motions) and the weakening gap
in distance between TWA and LCC, we believe that TWA is likely a near
edge of a larger population of stars (i.e., LCC). As more sensitive
data become available in future, namely next generation parallax 
measurements, one may find that the distribution of $\sim$10\,Myr old 
stars extends from TWA to all the way to LCC.

%

We note that we base our age estimate on Li-strong stars in the
relative age-dating of TWA, LCC/UCL, and $\beta$~Pictoris moving
group members.  One might therefore question whether the possible
existence and non-inclusion of Li-weak {\it true} members might
vitiate the validity of such a comparison.  Currently, in the
absence of accurate trigonometric parallax, there is no effective
way to identify Li-depleted members of young dispersed moving
groups.  This means that the same possible bias introduced by
including only Li-strong members exists equally in the TWA,
LCC/UCL, and the $\beta$~Pictoris Moving Group.  Therefore,
comparing the upper envelopes of Li\,$\lambda6708$ strength
distributions among young stellar groups should be a perfectly
valid method of relative age-dating.

\subsection{Comparison between HRD and CMD/Li ages}

\begin{figure}
{\centering \begin{tabular}{c}
\subfigure[HRD]{\resizebox*{0.65\columnwidth}{!}{\includegraphics{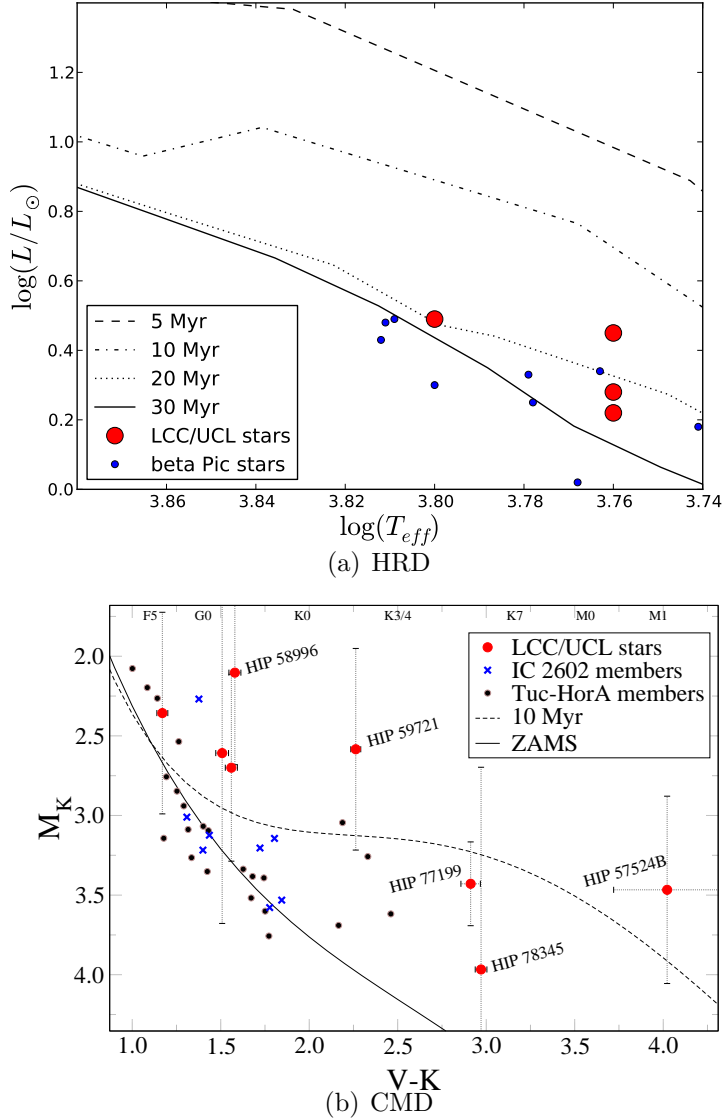}}} \\
\subfigure[CMD]{\resizebox*{0.55\columnwidth}{!}{\includegraphics{f3b.eps}}} \\
\end{tabular}\par}

\caption{(Top) LCC/UCL stars plotted on a Hertzsprung-Russell diagram (HRD)
along with theoretical isochrones from \citet{Siess}. Several F-type
$\beta$ Pictoris moving group (BPMG) members are located around the $\sim$30\,Myr
theoretical isochrone while the trusted age of the BPMG is based on various techniques
that all point to an age of $\sim$12\,Myr. This implies a significant
systematic discrepancy between HRD ages and ages obtained by other
methods. (Bottom) Same
LCC/UCL Hipparcos stars plotted on a color-magnitude diagram which does
not involve any theoretical models. When compared to empirical
$\sim$10\,Myr isochrones (from $\eta$ Cha, TWA, and BPMG members) and other slightly older ($\sim$30\,Myr) stars from
Tucana-HorA and IC\,2602, LCC/UCL stars appear to be $\sim$10\,Myr in
the observational domain. Distances are obtained from the reanalyzed
Hipparcos data \citep{Leeuwen}. The distance to HIP\,59721 is adapted
from its co-moving companion (HIP\,59716; $\pi=10.36\pm1.31$\,mas/yr) due
to the large error ($\pi=7.56\pm5.84$\,mas/yr) for HIP\,59721. For IC\,2602, $(m-M)_0=5.95$
was used following \citet{IC2602}.
}
\end{figure}

Among several commonly used age-dating methods -- position on a
color-magnitude diagram (CMD) or Hertzsprung-Russel diagram (HRD), stellar rotation,
Li\,$\lambda$6708 absorption strength, X-ray brightness, H$\alpha$ emission
strength, Galactic space motion, CaII\,HK index, IR excess emission -- the CMD, HRD,
and Li methods can provide quantitative age estimates for stars
in the 5-30 Myr age range. We already demonstrate the Li age in the
previous section. Using the CMD/HRD age-dating method requires a precise
distance to a star, and eight stars in Table 1 have measured trigonometric
parallaxes from Hipprcos \citep{Leeuwen}. Using the transformation
scheme from colors to effective temperatures and bolometric correction
values from \citep{Mamajek}, we plot four LCC/UCL Hipparcos stars on a HRD (Figure
3, left panel). Hipparcos stars from Table 1 sit on the theoretical 20 Myr isochrone
from \citet{Siess}, at first glance apparently in support of an age of $\sim$20\,Myr as
deduced by \citet{Mamajek}. However, as may be seen, various F-type $\beta$
Pictoris moving group members sit on the $\sim$30\,Myr isochrone which
is inconsistent with the age of the group ($\sim$12\,Myr). It implies an
age calibration problem between these two age-dating methods similar to
the case of inconsistent ages from CaII\,HK and Li ages \citep{CaII}.

In Figure 3, bottom panel, we plot eight Hipparcos Table 1 stars on a $V-K$ versus $M_K$
diagram (i.e., model-independent) along with an empirical 10\,Myr
isochrone from \citet{ABDor} and several dozen $\sim$30\,Myr old stars
from Tucana-Horologium Association and IC\,2602.  A useful
comparison would be plotting $\sim$20\,Myr old F/G-type stars but there
are no such suitable stars with reliably determined ages in the solar
neighborhood. As shown in the CMD, all eight
Hipparcos LCC/UCL stars are located on or above the 10\,Myr isochrone as
defined by $\eta$ Cha, TWA, and the $\beta$ Pictoris moving group (see
\citet{ABDor} for more details on the 10\,Myr empirical isochrone). This
is a corroborant demonstration that LCC/UCL stars are as young as stars in
$\eta$ Cha, TWA, and $\beta$ Pictoris moving group, and the claimed
older age of LCC/UCL is likely due either to a lack of or incorrect calibration of HRD ages
against empirical ages. Likewise, the relatively old age for the Upper-Sco region
recently deduced by \citet{Pecaut} requires additional scrutiny because
it is based on the same HRD age-dating method.

\section{Summary}

We spectroscopically identified $\sim\!\!100$ G/K/M type ScoCen members,
mostly LCC and UCL members, that show strong Li\,$\lambda$6708
absorption and/or H$\alpha$ emission features.  Comparison of Li absorption
strengths against those of other young stellar groups on a $V-K$ versus
lithium strength diagram indicates that the age of LCC/UCL is
$\sim\!\!10$\,Myr. Specifically, LCC/UCL stars must be younger than
stars in the $\beta$~Pictoris moving group whose age of 12 Myr has been
derived previously from kinematic traceback analysis. Based on plots of LCC/UCL
Hipparcos stars in color-magnitude and
Hertzsprung-Russel diagrams, we find that ages derived from the HR-diagram are systematically
older than CMD and Li ages; the HR-diagram ages are model-dependent whereas CMD and Li ages are 
primarily empirically anchored.  This difference can explain the discrepancy
between our young age and previously claimed older ages of UCL/LCC.
Because of the importance of accurate ages in many astrophysical
phenomena, a thorough cross-calibration of various age-dating methods
for young stars is in urgent need.

\acknowledgements{This research has made use of the SIMBAD \& Vizier databases,
operated at CDS, Strasbourg, France. We thank the anonymous referee for 
useful and constructive comments.  This research was partially supported by NASA grants to the University
of Georgia and UCLA.}


\newpage
\begin{landscape}
\small
\def\mc{\multicolumn}
\def\ch{\colhead}
\def\nd{--}
\def\noda{\mc{2}{c}{--}}
\setlength{\tabcolsep}{0.02in}
\begin{longtable}{c l c c r c c r@{$.$}l r@{$.$}l c c c r@{$\pm$}l r@{,}r@{,}r l}
\caption{Identified Members of Sco-Cen} \\
\hline\hline
{No.} & {Name}& {R.A.} &{Dec.}  & {Dist.} & {N} & \mc{3}{c}{Eq. Width} &
\mc{2}{c}{$V$} & {$V-K$}  & {$B-V$} & {$f$} & \mc{2}{c}{Rad. Vel.} & \mc{3}{c}{ $(U,V,W)$        } &{Note}

 \\ \cline{7-9}

  &      & \mc{2}{c}{(J2000)}    & {(pc)} & {}   & {Li}   & \mc{2}{c}{H$\alpha$} &
\mc{2}{c}{(mag)}& {(mag)} & {(mag)} &          & \mc{2}{c}{(km/sec)} & \mc{3}{c}{(km/sec)} &  \\
\hline
\endfirsthead
\caption{continued.}\\
\hline\hline
{No.} & {Name}& {R.A.} &{Dec.}  & {Dist.} & {N} & \mc{3}{c}{Eq. Width} &
\mc{2}{c}{$V$} & {$V-K$}  & {$B-V$} & {$f$} & \mc{2}{c}{Rad. Vel.} & \mc{3}{c}{ $(U,V,W)$        } &{Note}

 \\ \cline{7-9}

  &      & \mc{2}{c}{(J2000)}    & {(pc)} & {}   & {Li}   & \mc{2}{c}{H$\alpha$} &
\mc{2}{c}{(mag)}& {(mag)} & {(mag)} &          & \mc{2}{c}{(km/sec)} & \mc{3}{c}{(km/sec)} &  \\
\hline
\endhead
\hline
\endfoot
\mc{20}{l}{Lower Centaurus Crux}                                                                                          \\
\cline{1-4}
  1 &SSS 1132-3019 & 11:32:18.38 &$-$30:19:51.5&  42& 1 &500&$  -6$&7& 14&20d& 5.32& 1.62& -3.51&   \noda    &  \mc{3}{c}{\nd} & phot dist=55, TWA 30 \\
  2 &HIP 57524     & 11:47:24.58 &$-$49:53:02.9&  92& 2 &190&$   0$&6&  9&07 & 1.56& 0.65& -3.29& 13.4 & 1.5 & -6.4&-18.7& -5.0& phot dist=86, TWA 19A \\
  3 &HIP 57524 B   & 11:47:20.64 &$-$49:53:04.2&  92& 1 &330&$  -1$&6& 12&30d& 4.13& 1.52& -3.29& 15.3 & 3.2 & -5.7&-20.4& -4.6& phot dist=63, TWA 19B \\
  4 &TYC 8631-0128 & 11:55:57.75 &$-$52:54:00.8& 107& 3 &360&$   0$&1& 11&00 & 2.61& 1.06& -3.50& 11.5 & 1.5 &-10.5&-18.5& -4.8&  \\
  5 &SSS 1159-4510 & 11:59:27.87 &$-$45:10:19.2&  57& 1 &420&$  -4$&6& 14&54p& 5.47& 1.65& -2.92&   \noda    &  \mc{3}{c}{\nd} &  \\
  6 &SSS 1205-5331 & 12:05:12.66 &$-$53:31:23.1& 102& 1 &270&$  -1$&9& 13&54d& 4.23& 1.53& -3.06&   \noda    &  \mc{3}{c}{\nd} & \\
  7 &HIP 58996     & 12:05:47.52 &$-$51:00:12.1& 110& 1 &240&$   0$&9&  8&89 & 1.58& 0.66& -3.56& 16.3 & 2.5 & -8.6&-24.0& -6.0& phot dist=78.0 \\
  8 &SSS 1208-5850 & 12:08:20.60 &$-$58:50:15.1&  79& 1 &200&$  -3$&4& 15&35d& 5.53& 1.67& -3.08&   \noda    &  \mc{3}{c}{\nd} &  \\
  9 &SSS 1210-4855 & 12:10:10.34 &$-$48:55:45.9& 104& 1 &360&$  -0$&3& 11&21p& 2.79& 1.14& -3.08&   \noda    &  \mc{3}{c}{\nd} & \\
 10 &TYC 8636-2515 & 12:12:35.79 &$-$55:20:27.2&  99& 4 &300&$   0$&2& 10&48 & 2.36& 0.95& -3.22& 14.7 & 1.5 & -6.1&-20.9& -6.7&  \\
 11 &TYC 8978-3494 & 12:12:48.93 &$-$62:30:31.9&  63& 1 &350&$   0$&4& 11&77 & 3.81& 1.38& -2.52&   \noda    &  \mc{3}{c}{\nd} &  FS 623 \\
 12 &TYC 8978-5124 & 12:13:57.02 &$-$62:55:12.6&  98& 1 &310&$  -0$&2& 11&40 & 3.01& 1.28& -3.03&   \noda    &  \mc{3}{c}{\nd} &  \\
 13 &TYC 8242-1324 & 12:14:34.12 &$-$51:10:12.4& 103& 1 &270&$   0$&4& 10&29 & 2.15& 0.87& -3.35&   \noda    &  \mc{3}{c}{\nd} &  \\
 14 &TYC 8242-1324B& 12:14:31.88 &$-$51:10:15.7&  84& 1 &310&$  -0$&5& 13&57d& 4.48& 1.54& -3.35&   \noda    &  \mc{3}{c}{\nd} &  \\
 15 &HIP 59716     & 12:14:50.76 &$-$55:47:23.4&  97& 1 &100&$   1$&6&  8&45 & 1.17& 0.45& -3.54& 13.0 & 7.0 & -8.1&-19.8& -4.6& phot dist=86 \\
 16 &HIP 59721     & 12:14:52.35 &$-$55:47:03.5&  97& 1 &330&$  -0$&7&  9&77 & 2.26& 0.79& -3.54& 17.6 & 1.7 & -5.8&-23.6& -2.9& phot dist=76 \\
 17 &TYC 8637-1558 & 12:16:01.20 &$-$56:14:06.9&  69& 1 &290&$   0$&2& 11&50 & 3.54& 1.40& -2.99&   \noda    &  \mc{3}{c}{\nd} & broad lines \\
 18 &SSS 1216-5055 & 12:16:17.01 &$-$50:55:26.3& 105& 1 &150&$  -3$&7& 14&69d& 4.83& 1.57& -2.14&   \noda    &  \mc{3}{c}{\nd} & FS 625 \\
 19 &TYC 8986-0497 & 12:16:30.10 &$-$67:11:47.7&  72& 3 &430&$  -0$&2& 11&10 & 3.25& 1.39& -3.19&  7.0 & 5.0 & -6.8&-11.5& -5.5&  \\
 20 &TYC 9231-1185 & 12:16:40.31 &$-$70:07:36.1&  92& 1 &325&$   0$&0& 10&73 & 2.64& 1.07& -3.49&   \noda    &  \mc{3}{c}{\nd} & \\
 21 &TYC 8641-2187 & 12:18:58.05 &$-$57:37:19.1&  68& 4 &340&$  -0$&1&  9&87 & 2.50& 1.01& -3.22&   \noda    &  \mc{3}{c}{\nd} & broad lines\\
 22 &TYC 8983-0098 & 12:19:21.68 &$-$64:54:10.3&  66& 4 &340&$   0$&2& 10&12 & 2.72& 1.11& -3.21& 15.0 & 1.5 & -2.3&-18.3& -5.5& \\
 23 &SSS 1219-5018 & 12:19:59.38 &$-$50:18:38.1& 165& 1 &260&$  -1$&0& 12&89p& 3.24& 1.38& -2.85&   \noda    &  \mc{3}{c}{\nd} & 2MASS binary?, $\sim2\farcs5$ NS \\
 24 &TYC 8983-0795 & 12:20:54.56 &$-$64:57:24.2&  98& 1 &400&$  -0$&9& 10&39 & 2.31& 0.93& -3.53&   \noda    &  \mc{3}{c}{\nd} &  \\
 25 &TYC 8983-0564 & 12:21:30.84 &$-$64:03:52.7&  56& 1 &380&$  -0$&8& 10&83 & 3.42& 1.40& -3.33&   \noda    &  \mc{3}{c}{\nd} &  \\
 26 &TYC 8238-1462 & 12:21:55.69 &$-$49:46:12.4&  99& 2 &355&$  -0$&4& 10&02 & 2.01& 0.83& -3.42& 13.8 & 2.4 & -7.4&-21.5& -5.3&  \\
 27 &TYC 8234-2856 & 12:22:04.32 &$-$48:41:24.8& 101& 1 &340&$   0$&5& 10&51 & 2.35& 0.94& -3.23& 13.1 & 2.3 & -5.1&-19.2& -4.0&  \\
 28 &SSS 1222-5739 & 12:22:28.84 &$-$57:39:12.2&  81& 1 &240&$  -4$&3& 14&59d& 5.09& 1.59& -2.81&   \noda    &  \mc{3}{c}{\nd} & \\
 29 &SSS 1222-6020 & 12:22:39.93 &$-$60:20:24.4& 106& 1 &140&$  -3$&0& 15&09d& 5.04& 1.58& -2.76&   \noda    &  \mc{3}{c}{\nd} & \\
 30 &SSS 1223-5540 & 12:23:14.33 &$-$55:40:16.1&  78& 1 &130&$ -11$&0& 14&85d& 5.27& 1.60& -2.47&   \noda    &  \mc{3}{c}{\nd} &  FS 632 \\
 31 &TYC 8983-0854 & 12:23:47.54 &$-$64:02:54.9& 101& 1 &375&$  -0$&7& 10&79 & 2.55& 1.03& -3.41&   \noda    &  \mc{3}{c}{\nd} &  \\
 32 &TYC 8979-1997 & 12:27:16.63 &$-$62:39:14.2&  91& 1 &375&$   0$&0& 10&90 & 2.78& 1.13& -3.15&   \noda    &  \mc{3}{c}{\nd} &  \\
 33 &TYC 8979-1683 & 12:28:25.44 &$-$63:20:58.6&  73& 3 &260&$   0$&6&  9&33 & 2.00& 0.83& -2.91& 13.9 & 1.9 & -3.0&-17.9& -5.1& \\
 34 &TYC 8654-2791 & 12:33:33.85 &$-$57:14:06.6& 101& 1 &345&$   0$&0& 10&89 & 2.62& 1.07& -3.23&   \noda    &  \mc{3}{c}{\nd} &  \\
 35 &TYC 8992-0605 & 12:36:39.02 &$-$63:44:43.4&  68& 3 &410&$   0$&3&  9&88 & 2.51& 1.01& -3.38&   \noda    &  \mc{3}{c}{\nd} & \\
 36 &TYC 8646-0166 & 12:36:59.00 &$-$54:12:17.9& 104& 1 &290&$   0$&6& 10&40 & 2.22& 0.89& -3.36&   \noda    &  \mc{3}{c}{\nd} &  \\
 37 &TYC 8658-1264 & 12:38:35.60 &$-$59:16:43.8& 123& 1 &380&$  -0$&6& 11&62 & 2.83& 1.15& -3.10&   \noda    &  \mc{3}{c}{\nd} & \\
 38 &SSS 1244-6902 & 12:44:14.57 &$-$69:02:35.4&  79& 1 &520&$  -3$&3& 13&34d& 4.43& 1.54& -2.64&   \noda    &  \mc{3}{c}{\nd} & FS 645 \\
 39 &TYC 8992-0420 & 12:44:34.85 &$-$63:31:46.1&  79& 2 &390&$  -0$&9& 10&79 & 2.91& 1.20& -3.05&   \noda    &  \mc{3}{c}{\nd} &  \\
 40 &TYC 8647-0324 & 12:45:48.85 &$-$54:10:58.3& 127& 1 &340&$  -0$&2& 11&28 & 2.54& 1.03& -3.23&   \noda    &  \mc{3}{c}{\nd} &  \\
 41 &TYC 9228-1355 & 12:47:21.99 &$-$68:08:40.0&  86& 1 &425&$  -0$&5& 10&88 & 2.85& 1.16& -3.50&   \noda    &  \mc{3}{c}{\nd} &  \\
 42 &SSS 1247-5050 & 12:47:35.99 &$-$50:50:51.9& 107& 1 &140&$  -4$&7& 14&98d& 4.97& 1.58& -2.98&   \noda    &  \mc{3}{c}{\nd} & \\
 43 &TYC 8651-0442 & 12:47:48.27 &$-$54:31:30.6&  75& 1 &460&$  -2$&2& 11&47 & 3.42& 1.40& -3.06&   \noda    &  \mc{3}{c}{\nd} &  \\
 44 &TYC 7783-1908 & 12:48:07.82 &$-$44:39:16.6&  76& 1 &260&$   0$&0&  9&73 & 2.22& 0.89& -3.11&   \noda    &  \mc{3}{c}{\nd} &  \\
 45 &TYC 8257-1545 & 12:50:51.44 &$-$51:56:35.4& 109& 1 &450&$  -2$&1& 11&68 & 3.05& 1.30& -3.35&   \noda    &  \mc{3}{c}{\nd} & 2MASS binary? $\sim3''$ EW \\
 46 &SSS 1251-5253 & 12:51:05.57 &$-$52:53:12.1& 102& 1 &395&$  -1$&1& 12&39d& 3.57& 1.41& -2.87&   \noda    &  \mc{3}{c}{\nd} & FS 650 \\
 47 &SSS 1251-5630 & 12:51:12.46 &$-$56:30:46.8&  70& 1 &460&$  -2$&1& 13&24d& 4.51& 1.54& -3.17&   \noda    &  \mc{3}{c}{\nd} &  \\
 48 &SSS 1252-5615 & 12:52:00.60 &$-$56:15:57.7&  93& 1 &340&$  -4$&7& 14&52d& 4.87& 1.57& -2.71&   \noda    &  \mc{3}{c}{\nd} & \\
 49 &SSS 1252-5553 & 12:52:14.72 &$-$55:53:37.2& 109& 1 &360&$  -1$&7& 12&54d& 3.57& 1.41& -3.12&   \noda    &  \mc{3}{c}{\nd} & FS 652 \\
 50 &SSS 1255-5355 & 12:55:55.95 &$-$53:55:31.1&  99& 1 &315&$  -4$&4& 14&50d& 4.80& 1.56& -2.90&   \noda    &  \mc{3}{c}{\nd} & \\
 51 &TYC 9245-0535 & 12:56:08.35 &$-$69:26:53.9&  68& 1 &430&$  -1$&9& 11&63 & 3.64& 1.41& -2.66&   \noda    &  \mc{3}{c}{\nd} &  FS 655 \\
 52 &TYC 8989-0583 & 12:56:09.46 &$-$61:27:25.3&  68& 3 &260&$  -0$&1&  9&45 & 2.19& 0.88& -2.93& 10.5 & 3.0 & -8.3&-18.3& -4.1&  \\
 53 &TYC 9245-0617 & 12:58:25.65 &$-$70:28:49.0&  75& 3 &350&$   0$&0&  9&92 & 2.38& 0.95& -3.32& 11.1 & 1.5 & -7.3&-17.1& -7.5&  \\
 54 &SSS 1259-6808 & 12:59:35.74 &$-$68:08:01.0&  59& 1 &540&$  -6$&3& 14&56d& 5.44& 1.65& -1.90&   \noda    &  \mc{3}{c}{\nd} & FS 658 \\
 55 &TYC 8648-0446 & 13:01:50.70 &$-$53:04:58.3& 136& 1 &290&$   0$&2& 11&09 & 2.30& 0.92& -3.28&   \noda    &  \mc{3}{c}{\nd} &  \\
 56 &TYC 8993-0409 & 13:02:47.06 &$-$62:13:58.9&  88& 1 &315&$   0$&8& 10&18 & 2.33& 0.93& -3.25&   \noda    &  \mc{3}{c}{\nd} & 2MASS binary $\sim9''$ NE \\
 57 &TYC 9242-0290 & 13:14:01.15 &$-$68:46:38.5& 106& 1 &360&$  -0$&4& 11&27 & 2.81& 1.14& -3.27&   \noda    &  \mc{3}{c}{\nd} & SB2?\\
 58 &TYC 8259-0689 & 13:14:23.86 &$-$50:54:01.8&  99& 1 &250&$   0$&5& 10&40 & 2.30& 0.92& -2.97&   \noda    &  \mc{3}{c}{\nd} &  \\
 59 &TYC 8674-2317 & 13:21:20.30 &$-$59:03:44.0&  73& 2 &430&$  -1$&2& 10&82 & 3.04& 1.30& -3.50& 15.5 & 1.8 & -1.3&-20.8& -4.4& \\
 60 &HIP 65423     & 13:24:35.15 &$-$55:57:24.0& 124& 1 &220&$   0$&9&  9&59 & 1.51& 0.63& -3.51&  8.1 & 2.1 & -8.9&-17.9& -4.3& phot dist=106.2 \\
 61 &TYC 8256-1840 & 13:27:05.98 &$-$48:56:17.9&  77& 1 &350&$  -0$&6& 10&69 & 2.88& 1.18& -3.12&   \noda    &  \mc{3}{c}{\nd} &  \\
 62 &TYC 7796-2110 & 13:34:31.92 &$-$42:09:30.5&  93& 1 &315&$  -0$&9& 10&70 & 2.61& 1.06& -3.10&   \noda    &  \mc{3}{c}{\nd} &  \\
 63 &TYC 7796-1788 & 13:37:57.32 &$-$41:34:41.7&  91& 1 &275&$   0$&6& 10&08 & 2.20& 0.88& -3.20&   \noda    &  \mc{3}{c}{\nd} &  \\
 64 &TYC 7800-0858 & 13:38:05.99 &$-$43:44:56.3& 114& 1 &310&$   0$&0& 11&14 & 2.60& 1.06& -3.49&   \noda    &  \mc{3}{c}{\nd} &  \\
 65 &TYC 7796-0286 & 13:38:49.37 &$-$42:37:23.4& 138& 1 &320&$   0$&1& 11&36 & 2.47& 0.99& -3.64&   \noda    &  \mc{3}{c}{\nd} &  \\
 66 &TYC 8261-1690 & 13:40:25.56 &$-$46:33:51.3& 102& 2 &340&$  -1$&5& 11&38 & 2.94& 1.23& -2.79& 10.1 & 1.9 & -2.1&-17.0& -2.7&  \\
 67 &TYC 8266-2914 & 13:44:24.45 &$-$47:06:33.9&  93& 4 &310&$   0$&5& 10&50 & 2.47& 0.99& -3.47&   \noda    &  \mc{3}{c}{\nd} &  \\
 68 &TYC 9012-1005 & 13:44:42.84 &$-$63:47:49.2&  70& 4 &370&$  -0$&4& 10&88 & 3.14& 1.37& -3.16& 18.0 & 1.0 & -1.2&-24.0& -4.1&  SB? \\
 69 &TYC 8274-0030 & 13:45:56.02 &$-$52:22:25.3& 114& 1 &340&$  -0$&1& 11&34 & 2.75& 1.12& -2.98&   \noda    &  \mc{3}{c}{\nd} &  \\
 70 &TYC 8267-2879 & 13:54:42.13 &$-$48:20:57.6& 129& 1 &260&$   0$&0& 11&07 & 2.37& 0.95& -3.24&   \noda    &  \mc{3}{c}{\nd} &  \\
 71 &TYC 8271-0864 & 13:56:34.69 &$-$49:07:14.5& 136& 4 &310&$  -0$&2& 11&14 & 2.33& 0.93& -3.10&  5.6 & 1.7 &-11.8&-20.7& -4.3&  \\
\cline{1-4}
\mc{20}{l}{Upper Centaurus Lpus}                                                                                          \\
\cline{1-4}
 72 &TYC 7818-0504 & 14:30:13.56 &$-$43:50:09.7&  72& 1 &340&$  -0$&7& 10&46 & 2.82& 1.15& -2.80&   \noda    &  \mc{3}{c}{\nd} &  \\
 73 &SSS 1450-3458 & 14:50:34.04 &$-$34:58:56.1& 116& 2 &370&$  -0$&7& 12&10p& 3.22& 1.38& -3.12&  1.3 & 1.6 &  \mc{3}{c}{\nd} & no proper motion data\\
 74 &TYC 7325-0465 & 15:24:32.37 &$-$36:52:02.5& 154& 1 &340&$   0$&2& 10&87 & 1.94& 0.81& -3.41&  4.0 & 1.5 & -5.0&-23.3& -4.6&  \\
 75 &SSS 1533-3917 & 15:33:40.48 &$-$39:17:47.7&  62& 2 &280&$  -7$&9& 14&22u& 5.20& 1.60& -3.51& -2.0 & 5.0 &-12.9&-34.7&-23.0&  \\
 76 &SSS 1539-3451 & 15:39:46.38 &$-$34:51:02.6&  83& 1 &385&$  -0$&2& 12&89d& 4.12& 1.52& -3.02&   \noda    &  \mc{3}{c}{\nd} & \\
 77 &HIP 77199     & 15:45:47.65 &$-$30:20:54.9&  40& 3 &400&$  -0$&8&  9&37 & 2.91& 1.20& -3.03& -5.7 & 1.5 &-10.1&-20.2& -7.4& phot dist=17.4 (binary?) \\
 78 &TYC 6782-0900 & 15:47:07.49 &$-$25:19:46.4&  92& 1 &450&$  -0$&4& 11&00 & 2.83& 1.15& -3.04& -4.5 & 1.5 & -8.2&-22.6& -5.8&  \\
 79 &TYC 7328-1706 & 15:49:02.72 &$-$31:02:53.6& 121& 1 &350&$  -0$&1& 10&85 & 2.30& 0.92& -3.28& -9.9 & 3.5 &-13.9&-14.8& -4.9&  \\
 80 &TYC 7846-1538 & 15:53:27.32 &$-$42:16:00.2&  51& 3 &190&$   0$&8&  7&86 & 1.52& 0.64& -3.45& -0.2 & 3.5 & -6.9&-16.6& -5.4&  \\
 81 &TYC 7846-0833 & 15:56:44.01 &$-$42:42:29.9&  78& 1 &495&$  -1$&9& 11&88 & 3.61& 1.41& -3.10&   \noda    &  \mc{3}{c}{\nd} &  \\
 82 &HIP 78345     & 15:59:49.53 &$-$36:28:27.5&  65& 1 &425&$  -0$&3& 11&00 & 2.97& 1.25& -3.20& -0.3 & 1.2 & -3.4&-13.4& -7.7& phot dist=79.0 \\
 83 &SSS 1603-4018 & 16:03:05.46 &$-$40:18:25.8& \nd& 1 &310&$ -60$&5&\nd&\nd& \nd & \nd & \nd  &   \noda    &  \mc{3}{c}{\nd} & EX Lup \\
 84 &SSS 1606-2036 & 16:06:31.70 &$-$20:36:23.2&  70& 1 &450&$  -1$&3& 13&47d& 4.64& 1.55& -3.04&   \noda    &  \mc{3}{c}{\nd} & FS 810 \\
 85 &TYC 7349-2447 & 16:35:22.41 &$-$33:28:52.2&  72& 1 &415&$  -0$&8& 11&66 & 3.58& 1.41& -2.94&   \noda    &  \mc{3}{c}{\nd} &  \\
 86 &SSS 1639-3920 & 16:39:47.32 &$-$39:20:40.5&  99& 1 &440&$  -2$&0& 14&08d& 4.56& 1.55& -2.39&   \noda    &  \mc{3}{c}{\nd} & FS 844 \\
 87 &SSS 1652-3359 & 16:52:10.87 &$-$33:59:33.3&  86& 2 &325&$  -1$&1& 11&98d& 3.56& 1.40& -3.08&   \noda    &  \mc{3}{c}{\nd} & \\
\cline{1-4}
\mc{20}{l}{Upper Scorpius and young stars in the vicinity}                                                                  \\
\cline{1-4}
 88 &TYC 6801-0186 & 16:14:59.19 &$-$27:50:22.7& 130& 1 &355&$   0$&2& 10&98 & 2.29& 0.92& -3.33& -2.0 & 1.5 & -3.6&-16.6& -6.1&  \\
 89 &TYC 6798-0544 & 16:25:19.26 &$-$24:26:52.5&  87& 1 &400&$   1$&1& 10&06 & 2.26& 0.91& -3.50& -2.4 & 1.5 & -3.6&-13.7& -1.9&  \\
 90 &TYC 7344-0788 & 16:26:57.65 &$-$30:32:27.7&  98& 2 &440&$  -0$&5& 11&68 & 3.19& 1.38& -3.09& -4.2 & 1.6 &  \mc{3}{c}{\nd} &  \\
 91 &TYC 7344-0788B& 16:26:57.00 &$-$30:32:23.3&  94& 2 &425&$  -1$&0& 12&46d& 3.72& 1.43& -3.09&   \noda    &  \mc{3}{c}{\nd} & \\
 92 &TYC 6816-0234 & 17:13:32.84 &$-$26:02:06.9& 101& 1 &350&$   0$&6& 10&14 & 2.08& 0.85& -3.17& -7.0 & 5.0 & -6.9&-19.5& -4.2&  \\
 93 &TYC 6820-0223 & 17:15:03.62 &$-$27:49:39.4&  59& 1 &580&$  -2$&3& 10&56 & 3.18& 1.38& -3.15& -1.1 & 1.8 & -0.9&-10.6& -6.8&  \\
 94 &HIP 84642     & 17:18:14.71 &$-$60:27:26.7&  59& 3 &185&$   0$&6&  9&51 & 1.98& 0.82& -3.42&  0.6 & 1.2 &-14.3&-26.4& -1.0& $\ddagger$, phot dist=65.3 \\
 95 &SSS 1719-4615 & 17:19:42.09 &$-$46:15:26.5&  35& 2 &520&$ -10$&2& 12&93p& 5.17& 1.59& -3.25&   \noda    &  \mc{3}{c}{\nd} & Wack3672, Flare star,  \\
 96 &SSS 1724-3914 & 17:24:53.51 &$-$39:14:43.8& 128& 2 &275&$   0$&0& 11&88p& 2.94& 1.23& -3.46& -3.2 & 1.9 &-12.6&-42.1&-32.4& \\
 97 &TYC 8728-2262 & 17:29:55.08 &$-$54:15:48.1&  72& 2 &310&$   0$&2&  9&54 & 2.18& 0.88& -3.21& -0.5 & 3.7 & -9.8&-17.2& -8.9&  \\
 98 &TYC 5672-0216 & 17:37:46.48 &$-$13:14:45.6&  45& 3 &260&$  -0$&8& 10&11 & 3.27& 1.39& -2.56&   \noda    &  \mc{3}{c}{\nd} & FS 903 \\
 99 &HIP 86598     & 17:41:49.04 &$-$50:43:27.5&  72& 2 &190&$   0$&9&  8&33 & 1.34& 0.51& -3.65&  1.7 & 1.7 & -7.0&-19.4&-10.5& phot dist=71.3 \\
100 &TYC 8742-2065 & 17:48:33.74 &$-$53:06:42.9&  55& 3 &260&$   0$&4&  8&99 & 2.21& 0.89& -3.15& -0.2 & 1.5 & -6.2&-12.0& -5.9&  \\
101 &SSS 1751-4854 & 17:51:34.16 &$-$48:54:55.4&  54& 2 &290&$  -3$&8& 13&14d& 4.76& 1.56& -2.93&   \noda    &  \mc{3}{c}{\nd} & \\
102 &SSS 1814-3246 & 18:14:22.09 &$-$32:46:10.8&  71& 1 &125&$  -1$&6& 12&79p& 4.25& 1.53& -2.65&   \noda    &  \mc{3}{c}{\nd} &  \\
103 &SSS 1818-3710 & 18:18:35.44 &$-$37:10:11.5&  60& 2 &330&$  -7$&1& 14&65d& 5.70& 1.71& -3.02&   \noda    &  \mc{3}{c}{\nd} &  \\
104 &TYC 7408-0054 & 18:50:44.47 &$-$31:47:46.8&  50& 2 &425&$  -1$&6& 11&31 & 3.85& 1.47& -3.09& -3.0 & 6.0 & -4.1&-16.3& -8.6&  \\
\end{longtable}
\begin{list}{-}{}
\item  For non-Hipparcos stars, distances are photometrically estimated based on an 
   empirical $\sim$10\,Myr isochrone on a $V-K$ versus $M_K$ diagram 
   (e.g., Fig. 2 of \citealt{ARAA}). A typical uncertainty is about 30\%.
\item  column `N' indicates the number of independent measurements
   (listed EW values for Li and H$\alpha$ are average of N measurements).
\item  Equivalent widths for Li\,$\lambda6708$ and H$\alpha$ are in m\AA\ and \AA\, 
   respectively. `$-$' sign indicates emission.
\item  suffix `d' after V indicates photometric data from DENIS
\item  suffix `p' after V indicates our own V-band photometry + 2MASS K
\item  suffix `u' after V indicates USNO Rmag + 2MASS K
\item  no suffix after V means Vmag come from either Hipparcos or Tycho-2 and K from 2MASS
\item  B-V colors are interpolated from V-K values using \citet{KH95}.
\item  X-ray data are from ROSAT All Sky Survey \citep{RASS,RASSFSC} and 
   $f\equiv\log L_X/L_{bol}$. For binaries, X-ray counts
   are divided according to each star's optical brightness.
\item  HIP 60913 may be a member (good kinematics and Li=215\,m\AA) 
   with low X-ray luminosity ($\log L_X/L_{bol}=-4.34$).
\item  HIP 76063 may be a member (A-type star located on zero-age main 
   sequence; \citealt{ARAA}).
\item  $\ddagger$ Based on $UVW$, HIP 84642 may instead be a Tucana/HorA member 
   (Table~7 of \citealt{ARAA}).
\item  `FS' designation indicates an X-ray variable star \citep{FS}.
\item  Some ($\sim$30) stars listed in the above table were previously 
   identified by \citet{Mamajek}.
\end{list}

\end{landscape}

\end{document}